\documentclass[twocolumn,12pt]{iopart}

\usepackage{graphicx}

\begin{document}

\title[]{Proposal for the complete high-dimensional Greenberger-Horne-Zeilinger state measurement}
\author{Zhi Zeng$^{1,2,^*}$}
\address{{$^1$Institute of Signal Processing and Transmission, Nanjing University of Posts and Telecommunications, Nanjing 210003, China}  \\
{$^2$Key Lab of Broadband Wireless Communication and Sensor Network Technology, Ministry of Education, Nanjing University of Posts and Telecommunications, Nanjing 210003, China}}
\eads{\mailto{zengzhiphy@yeah.net}}
\vspace{10pt}

\begin{abstract}
A theoretical proposal for the complete analysis of high-dimensional Greenberger-Horne-Zeilinger (GHZ) state is presented in this Letter. We first demonstrate the approach for the complete three-photon GHZ state measurement in three dimensions, and then generalize it to the situation of $N$-photon system in $d$ dimensions. In our approach, the photonic hyperentanglement and quantum Fourier transform are both utilized. The presented proposal will be useful for the high-dimensional multi-photon quantum computation and quantum communication.
\end{abstract}

Quantum entanglement has been widely researched in the past years, and it can benefit the quantum science and technology a lot \cite{a1,a2,a3,a4,a5,a6,a7,a8,a9,a10,a11,a12,a13,a14}. In many high-dimensional quantum information processing tasks, the discrimination of a set of orthogonal high-dimensional entangled states is often required for getting the information \cite{1}. However, when the discrimination process is not complete, the information capacity and communication efficiency will be both decreased \cite{2,3,a15,a16}. It has been shown that the complete analysis of the mutually orthogonal entangled states is impossible, if just the linear optics is utilized \cite{4}. A variety of degrees of freedom (DOFs) of photon system can be used for encoding the high-dimensional quantum information, such as the spatial-mode DOF \cite{5}, orbital angular momentum (OAM) DOF \cite{6}, optical frequency DOF \cite{7} and time-bin DOF \cite{8}. In the multi-qudit quantum computation and quantum communication, the high-dimensional Greenberger-Horne-Zeilinger (GHZ) state is an essential resource, the preparation of which has been explored in both the theory and experiment \cite{9,10,11,12}. In 2018, Erhard \emph{et al.} have experimentally created a three-particle GHZ state entangled in three levels for every particle, which can carry more information than entangled states of qubits \cite{9}. In 2021, Paesani \emph{et al.} showed how to generate GHZ states in arbitrary dimensions and numbers of photons, resorting to the linear optical circuits described by Fourier transform matrices \cite{10}. In 2022, Bell \emph{et al.} presented a protocol for the near-deterministic generation of $N$-photon, $d$-dimensional photonic GHZ states using an array of $d$ non-interacting single-photon emitters \cite{11}. Recently, the experimental demonstration and certification of a high-dimensional GHZ state in a superconducting quantum processor was also reported \cite{12}.

The high-dimensional GHZ state measurement (HDGSM) is quite important and useful for the qudit-based quantum information technology. But up to now, there is no efficient proposal for the complete HDGSM of photon system. In this Letter, we propose a simple and efficient approach for the photonic complete HDGSM by using the hyperentanglement and quantum Fourier transform (QFT), which are employed for obtaining the parity information and relative phase information of high-dimensional GHZ states, respectively. First, we demonstrate the complete HDGSM for the three-dimensional three-photon system in spatial-mode DOF, resorting to the auxiliary entanglement in OAM DOF. Then, we generalize this approach to the complete HDGSM for $d$-dimensional $N$-photon system. Finally, a brief discussion on the feasibility of the presented proposal is given.

The 27 orthogonal three-dimensional GHZ states of three-photon system in spatial-mode DOF can be written as
\begin{eqnarray}
|\psi_{mn}^{k}\rangle = \frac{1}{\sqrt{3}}\sum_{j=0}^{2}e^{2i\pi jk/3}|j\rangle \otimes |j \oplus m\rangle \otimes |j \oplus n\rangle_{ABC},
\end{eqnarray}
where $m,n,k,j=0,1,2$. Here, $j \oplus m \equiv (j+m) {\rm{mod}} D$, and $D$ is the dimension of entanglement. $A$, $B$ and $C$ are the three entangled photons, and $|0\rangle$, $|1\rangle$ and $|2\rangle$ are the three different spatial-modes of photon. In the above expression, $m$ and $n$ represent the parity information of high-dimensional GHZ states, and $k$ represents the relative phase information of high-dimensional GHZ states. In order to distinguish the 27 three-dimensional GHZ states completely, all of the $m$, $n$ and $k$ should be determined. In our approach, $m$ and $n$ are determined by using the auxiliary high-dimensional GHZ state in OAM DOF, and $k$ is determined by using the QFT on photons.

The three-dimensional hyperentangled GHZ state of three-photon system in both spatial-mode and OAM DOFs can be written as
\begin{eqnarray}
|\Psi\rangle_{ABC} = |\psi_{mn}^{k}\rangle \otimes |\phi_{00}^{0}\rangle.
\end{eqnarray}
$|\phi_{00}^{0}\rangle$ is the auxiliary entanglement in OAM DOF we utilized, the form of which is
\begin{eqnarray}
|\phi_{00}^{0}\rangle = \frac{1}{\sqrt{3}}(|aaa\rangle + |bbb\rangle + |ccc\rangle)_{ABC}.
\end{eqnarray}
Here, $|a\rangle$, $|b\rangle$, and $|c\rangle$ are the three different levels of OAM entanglement of photons $A$, $B$ and $C$.

To accomplish the complete analysis of three-dimensional GHZ states in spatial-mode DOF, the three-photon system is initially prepared in the hyperentanglement $|\psi_{mn}^{k}\rangle\otimes|\phi_{00}^{0}\rangle$. Photons $A$, $B$ and $C$ are injected into the same setups, and one of the schematic diagram is shown in Fig. 1. After the path control on the three photons, the hyperentanglement evolves as
\begin{eqnarray}
|\psi_{00}^{k}\rangle\otimes|\phi_{00}^{0}\rangle \rightarrow |\psi_{00}^{k}\rangle \otimes \frac{1}{\sqrt{3}} (|aaa\rangle + |bbb\rangle + |ccc\rangle)_{ABC},  \nonumber  \\
|\psi_{01}^{k}\rangle\otimes|\phi_{00}^{0}\rangle \rightarrow |\psi_{01}^{k}\rangle \otimes \frac{1}{\sqrt{3}} (|aab\rangle + |bbc\rangle + |cca\rangle)_{ABC},  \nonumber  \\
|\psi_{02}^{k}\rangle\otimes|\phi_{00}^{0}\rangle \rightarrow |\psi_{02}^{k}\rangle \otimes \frac{1}{\sqrt{3}} (|aac\rangle + |bba\rangle + |ccb\rangle)_{ABC},  \nonumber  \\
|\psi_{10}^{k}\rangle\otimes|\phi_{00}^{0}\rangle \rightarrow |\psi_{10}^{k}\rangle \otimes \frac{1}{\sqrt{3}} (|aba\rangle + |bcb\rangle + |cac\rangle)_{ABC},  \nonumber  \\
|\psi_{11}^{k}\rangle\otimes|\phi_{00}^{0}\rangle \rightarrow |\psi_{11}^{k}\rangle \otimes \frac{1}{\sqrt{3}} (|abb\rangle + |bcc\rangle + |caa\rangle)_{ABC},  \nonumber  \\
|\psi_{12}^{k}\rangle\otimes|\phi_{00}^{0}\rangle \rightarrow |\psi_{12}^{k}\rangle \otimes \frac{1}{\sqrt{3}} (|abc\rangle + |bca\rangle + |cab\rangle)_{ABC},  \nonumber  \\
|\psi_{20}^{k}\rangle\otimes|\phi_{00}^{0}\rangle \rightarrow |\psi_{20}^{k}\rangle \otimes \frac{1}{\sqrt{3}} (|aca\rangle + |bab\rangle + |cbc\rangle)_{ABC},  \nonumber  \\
|\psi_{21}^{k}\rangle\otimes|\phi_{00}^{0}\rangle \rightarrow |\psi_{21}^{k}\rangle \otimes \frac{1}{\sqrt{3}} (|acb\rangle + |bac\rangle + |cba\rangle)_{ABC},  \nonumber  \\
|\psi_{22}^{k}\rangle\otimes|\phi_{00}^{0}\rangle \rightarrow |\psi_{22}^{k}\rangle \otimes \frac{1}{\sqrt{3}} (|acc\rangle + |baa\rangle + |cbb\rangle)_{ABC}.
\end{eqnarray}

\begin{figure}
\centering
\includegraphics*[width=0.65\textwidth]{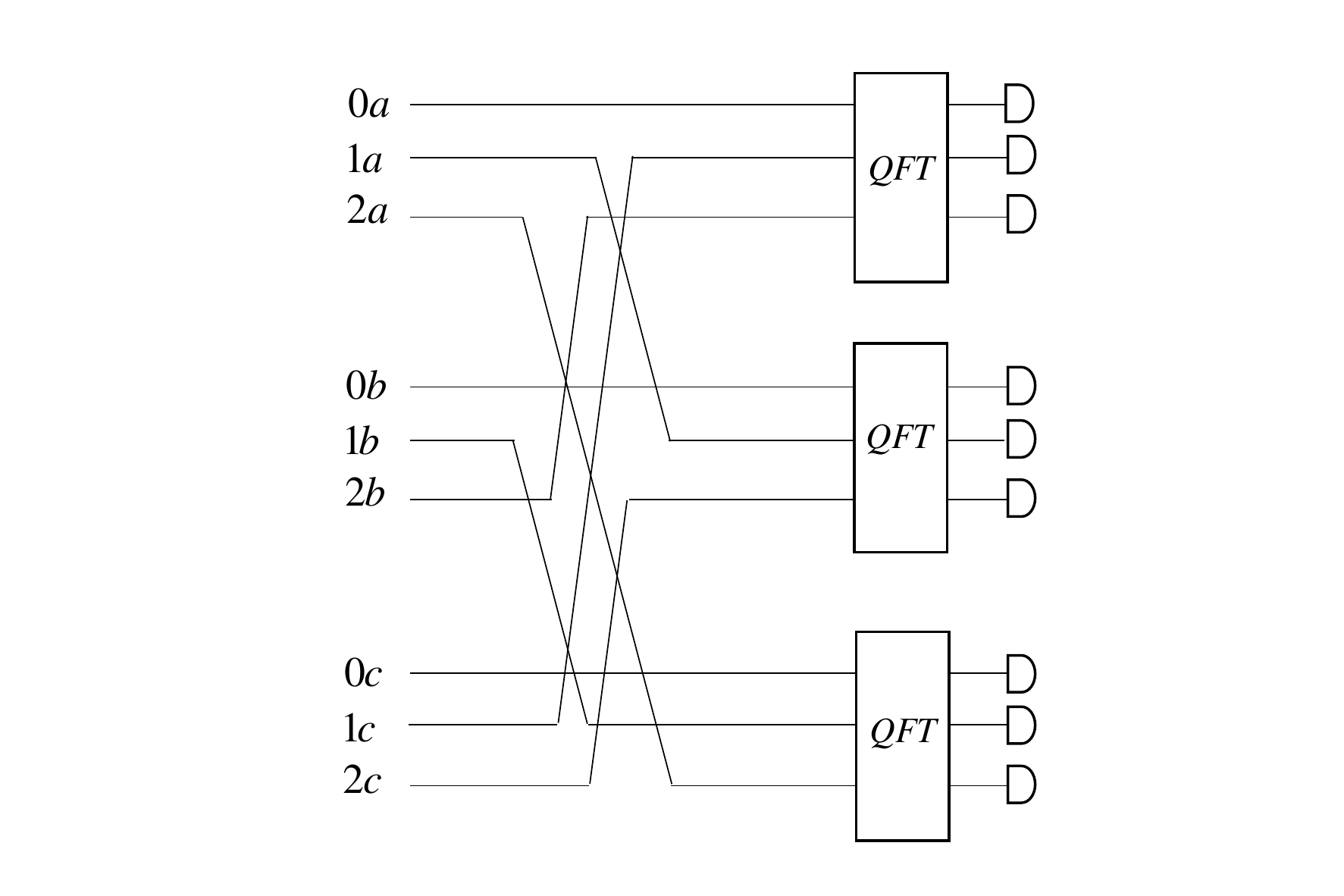}
\caption{Schematic diagram of our complete HDGSM proposal for the three-dimensional three-photon system. The path control in photonic hyperentanglement and three-dimensional QFT are used for obtaining the parity information and relative phase information of the 27 three-dimensional GHZ states, respectively. With the detection results of single photon detectors, the complete HDGSM can be accomplished.}
\end{figure}

It is easy to find that the three-dimensional GHZ states in spatial-mode DOF are invariant during the operation, and their parity information $m,n$ can be obtained through the detection on OAM entanglement, as shown in Table 1.

\begin{table}
\centering\caption{Corresponding relations between the initial states and the possible detections in OAM DOF.}
\begin{tabular}{cc ccccccccccc}
\hline
Initial states & & & & Possible detections in OAM DOF \\
\hline
$|\psi_{00}^{k}\rangle$ &&&& $aaa$, $bbb$, $ccc$.   \\
$|\psi_{01}^{k}\rangle$ &&&& $aab$, $bbc$, $cca$.   \\
$|\psi_{02}^{k}\rangle$ &&&& $aac$, $bba$, $ccb$.   \\
$|\psi_{10}^{k}\rangle$ &&&& $aba$, $bcb$, $cac$.   \\
$|\psi_{11}^{k}\rangle$ &&&& $abb$, $bcc$, $caa$.   \\
$|\psi_{12}^{k}\rangle$ &&&& $abc$, $bca$, $cab$.   \\
$|\psi_{20}^{k}\rangle$ &&&& $aca$, $bab$, $cbc$.   \\
$|\psi_{21}^{k}\rangle$ &&&& $acb$, $bac$, $cba$.   \\
$|\psi_{22}^{k}\rangle$ &&&& $acc$, $baa$, $cbb$.   \\

\hline
\end{tabular}
\end{table} 

With the deterministic $m$ and $n$, there are three states with different relative phase information $k$, and they can be discriminated with the help of three-dimensional QFT on the spatial-mode entanglement, which can be expressed as 
\begin{eqnarray}
|0\rangle &\rightarrow& \frac{1}{\sqrt{3}}(|0\rangle + |1\rangle + |2\rangle),\nonumber  \\
|1\rangle &\rightarrow& \frac{1}{\sqrt{3}}(|0\rangle + e^{2i\pi/3}|1\rangle + e^{4i\pi/3}|2\rangle),\nonumber  \\
|2\rangle &\rightarrow& \frac{1}{\sqrt{3}}(|0\rangle + e^{4i\pi/3}|1\rangle + e^{2i\pi/3}|2\rangle).
\end{eqnarray}
Here we take $m,n=0$ as an example to illustrate, and the evolution of the three spatial-mode GHZ states ($k=0,1,2$) after three-dimensional QFT is 
\begin{eqnarray}
|\psi_{00}^{0}\rangle \rightarrow &&\frac{1}{3}(|000\rangle + |012\rangle + |021\rangle + |102\rangle+ |111\rangle \nonumber  \\ &&+ |120\rangle+ |201\rangle+ |210\rangle+ |222\rangle)_{ABC},\nonumber  \\
|\psi_{00}^{1}\rangle \rightarrow &&\frac{1}{3}(|002\rangle + |011\rangle + |020\rangle + |101\rangle+ |110\rangle \nonumber  \\ &&+ |122\rangle+ |200\rangle+ |212\rangle+ |221\rangle)_{ABC},\nonumber  \\
|\psi_{00}^{2}\rangle \rightarrow &&\frac{1}{3}(|001\rangle + |010\rangle + |022\rangle + |100\rangle+ |112\rangle \nonumber  \\ &&+ |121\rangle+ |202\rangle+ |211\rangle+ |220\rangle)_{ABC}.
\end{eqnarray}
The above three states can be completely distinguished by detecting the three-dimensional entanglement in spatial-mode DOF, and it should be noted that this process is also suitable for other $m,n$. That is to say, the relative phase information $k$ can be obtained by using the three-dimensional QFT, as shown in Table 2. With Table 1 and Table 2, all of the 27 three-dimensional three-photon GHZ states in spatial-mode DOF can be distinguished.

\begin{table}
\centering\caption{Corresponding relations between the initial states and the possible detections in spatial-mode DOF.}
\begin{tabular}{cc ccccccccccc}
\hline
Initial states & & & & Possible detections in spatial-mode DOF\\
\hline
$|\psi_{mn}^{0}\rangle$ &&&& $000$, $012$, $021$, $102$, $111$, $120$, $201$, $210$, $222$.   \\
$|\psi_{mn}^{1}\rangle$ &&&& $002$, $011$, $020$, $101$, $110$. $122$, $200$, $212$, $221$.  \\
$|\psi_{mn}^{2}\rangle$ &&&& $001$, $010$, $022$, $100$, $112$, $121$, $202$, $211$, $220$.   \\

\hline
\end{tabular}
\end{table} 

By using the $d$-dimensional $N$-photon OAM GHZ state as the auxiliary entanglement and the $d$-dimensional QFT on spatial-mode DOF, our approach can be extended to the complete $N$-photon HDGSM in $d$ dimensions. The $d^{N}$ orthogonal $d$-dimensional GHZ states of $N$-photon system can be written as
\begin{eqnarray}
|\psi_{x_1x_2\cdots x_{N-1}}^{k}\rangle = \frac{1}{\sqrt{d}}\sum_{j=0}^{d-1}e^{2i\pi jk/d}|j\rangle |j \oplus x_1\rangle |j \oplus x_2\rangle \cdots |j \oplus x_{N-1}\rangle_{AB\cdots N},
\end{eqnarray}
where $x_1,x_2,\cdots, x_{N-1},k,j=0,1,2,\cdots,d-1$, and $|0\rangle$, $|1\rangle, |2\rangle, \cdots, |d-1\rangle$ are the $d$ different spatial-modes of photon. Here, $x_1,x_2,\cdots, x_{N-1}$ represent the parity information, and $k$ represents the relative phase information.

To get the parity information of high-dimensional entanglement, the following auxiliary GHZ state in OAM DOF is required,
\begin{eqnarray}
|\phi_{00\cdots 0}^{0}\rangle = \frac{1}{\sqrt{d}}\sum_{j=0}^{d-1} |j\rangle |j\rangle \cdots |j\rangle_{AB\cdots N},
\end{eqnarray}
where $j=0,1,2,\cdots,d-1$ represent the $d$ different levels of OAM entanglement. The effect of path control on a single photon between two different DOFs can be expressed as
\begin{eqnarray}
|j_1\rangle^{S} \otimes |j_2\rangle^{O}_{A(B,\cdots,N)} \rightarrow |j_1\rangle^{S} \otimes |j_1 \oplus j_2\rangle^{O}_{A(B,\cdots,N)}.
\end{eqnarray}
Here $j_1,j_2=0,1,2,\cdots,d-1$, and $|j_1\rangle^{S}$ and $|j_2\rangle^{O}$ represent the states in spatial-mode DOF and OAM DOF, respectively. After the path control in high-dimensional multi-photon system, the hyperentanglement in two DOFs evolves as
\begin{eqnarray}
&&|\psi_{x_1x_2\cdots x_{N-1}}^{k}\rangle \otimes |\phi_{00\cdots 0}^{0}\rangle \nonumber  \\ &&\rightarrow |\psi_{x_1x_2\cdots x_{N-1}}^{k}\rangle \otimes \frac{1}{\sqrt{d}}\sum_{j=0}^{d-1} |j\rangle |j \oplus x_1\rangle |j \oplus x_2\rangle \cdots |j \oplus x_{N-1}\rangle_{AB\cdots N}.
\end{eqnarray}
The high-dimensional GHZ states in spatial-mode DOF are invariant during the operation, and their parity information $x_1,x_2,\cdots, x_{N-1}$ can be identified by the detection results on photonic OAM enatnglement, which correspond to $|j\rangle |j \oplus x_1\rangle |j \oplus x_2\rangle \cdots |j \oplus x_{N-1}\rangle_{AB\cdots N}$. 

The relative phase information of high-dimensional entanglement can be obtained by using the $d$-dimensional QFT, which can be expressed as
\begin{eqnarray}
{\rm{QFT}}|z\rangle = \frac{1}{\sqrt d}\sum_{j=0}^{d-1}e^{2i\pi zj/d}|j\rangle,
\end{eqnarray}
where $z=0,1,2,\cdots,d-1$. After the operation of QFT on the $N$ photons in spatial-mode DOF, the high-dimensional entanglement evolves as 
\begin{eqnarray}
{\rm{QFT}} |\psi_{x_1x_2\cdots x_{N-1}}^{k}\rangle  = \frac{1}{\sqrt d}\sum_{j=0}^{d-1}u_{jx_1x_2\cdots x_{N-1}}^{k}|j\rangle \otimes |y_1\rangle|y_2\rangle \cdots |y_{N-1}\rangle_{AB\cdots N}.
\end{eqnarray}
In the above equation, $u_{jx_1x_2\cdots x_{N-1}}^{k}$ denotes the phase factor that will not influence the final results, and $y_1, y_2, \cdots, y_{N-1}$ satisfy
\begin{eqnarray}
y_1 \oplus y_2 \oplus \cdots \oplus y_{N-1} \oplus j \oplus k = 0,
\end{eqnarray}
where $y_1, y_2, \cdots, y_{N-1}=0,1,2,\cdots,d-1$. Thus, the relative phase information $k$ can be determined, which corresponds to the detection results $|j\rangle|y_1\rangle|y_2\rangle \cdots |y_{N-1}\rangle_{AB\cdots N}$. When both of the parity information and relative phase information have been identified, the complete HDGSM for $d$-dimensional $N$-photon system is accomplished.

In this Letter, we have presented the proposal for the complete HDGSM of multi-photon system in arbitrary dimensions, resorting to the photonic hyperentanglement and QFT. However, we should acknowledge that this proposal is still theoretical, which is not easy to realize with the current technology. On the one hand, so far, the experimental preparation of high-dimensional hyperentangled GHZ state remains a challenge. Fortunately, both of the two-dimensional multi-photon hyperentangled GHZ state and high-dimensional two-photon hyperentangled state have already been realized in experiment \cite{13,14,15,16,17}. For example, in 2018, Wang \emph{et al.} experimentally demonstrated an 18-qubit GHZ entanglement by simultaneous exploiting three different DOFs of six photons, including their paths, polarization, and OAM \cite{14}. In 2022, Hu \emph{et al.} have experimentally generated a high-dimensional hyperentangled photon pairs from two parallel spontaneous four-wave mixing processes with beam displacer interferometers \cite{17}. Therefore, we are confident that the high-dimensional hyperentangled GHZ state of multi-photon system will be efficiently prepared with the development of quantum technology. In our scheme, the spatial-mode DOF and OAM DOF of photon system are treated as the target qudit and auxiliary qudit, respectively. The quantum manipulations of these two different DOFs are quite mature \cite{a17,a18,a19,a20}, for example, Kong \emph{et al.} experimentally demonstrated the manipulation of eight-dimensional Bell-like states based on photonic spin-OAM hyperentanglement in 2019 \cite{a18}. In 2021, Hiekkamäki \emph{et al.} studied the two-photon interference in multiple transverse-spatial modes along a single beam-path, and observed the coalescence and anticoalescence in different three- and four-dimensional spatial-mode multiports \cite{a20}. Actually, some other DOFs of photon (including the frequency qudit and time-bin qudit) may also be utilized to get the parity information of high-dimensional GHZ state in spatial-mode DOF. On the other hand, the high-dimensional QFT on photon is required in our proposal \cite{18,19,20}, which is exploited to discriminate the relative phase information of target high-dimensional state, as shown in Eqs. (6) and (12). In 2022, Shi \emph{et al.} proposed a photonic QFT protocol, in which a single atom-coupled cavity system implements the photon-photon interactions and no active feedforward control is needed \cite{19}. Moreover, they also showed that a QFT with tens of photons may be possible with state-of-the-art cavity-QED systems \cite{19}. In conclusion, to our knowledge, the presented proposal is the first complete HDGSM scheme for entangled photons. Although it is not easy to accomplish in experiment at the present time, it is an important step toward the complete analysis of high-dimensional GHZ state. We hope our proposal can be realized in the near future, and it will have useful applications in the high-dimensional quantum computation and quantum communication. 

\section*{Disclosures}
The author declares no conflicts of interest.

\section*{Data availability} 
Data underlying the results presented in this paper are not publicly available at this time but may be obtained from the author upon reasonable request.

\end{document}